# Design of Three-Phase Hybrid Active Power Filter for Compensating the Harmonic Currents of Three-Phase System

ZUBAIR AHMED MEMON*, MOHAMMAD ASLAM UQAILI**, AND MUKHTIAR AHMED UNAR***

## ABSTRACT

Power quality standards (IEEE-519) require to limit the total harmonic distortion within satisfactory range caused by power electronic based devices. Our work deals with the design of hybrid active filter to reduce current perturbations produced by power electronics based devices. The Instantaneous Active and Reactive Power Method (pq) is used to perform the identification of disturbing currents. The pq algorithm creates a reference current, whereas, this reference current is tracked by the current of the voltage source converter. The currents of the voltage source converter are controlled by hysteresis controller. Simulation results showed that the hybrid active filter can compensate the harmonic currents effectively and improve power quality.

## 1.    INTRODUCTION

With the development of modern industrial technology a large number of non-linear loads are used in power systems which cause harmonic pollution. At the same time the power quality and safe operation becomes worse sharply. Therefore, mitigation of harmonics is essential under the situation. In the literature, numerous topologies of active filters have been presented for mitigation of harmonics [1-3]. The APFs (Active Power Filters) topologies are not cost-effective for the application of high power because of their high rating and very high switching frequency of PMW (Pulse Width Modulator) converter. Thus LC PPFs (Passive Power Filters) are used for harmonic filtration of such large nonlinear loads.

LC and high reliability is the main advantage of passive filters [4]. However, passive filters suffer from some shortcomings for example, the performance of these filters is affected due to the varying impedance of the system and with the utility system the series and parallel resonances may be created, which cause current harmonics increase in the supply [5-6]. As discussed above both active filters and passive filters suffer from a number of disadvantages. Therefore, another solution of harmonic mitigation, called HAPF (Hybrid Active Power Filter), has been introduced. HAPF provides the combined advantages of APF and PPF and eliminate their disadvantages. These topologies are cost effective solutions of the high-power power quality problems with well filtering performance. This research paper is restricted to the (HAPF). The (HAPF) is specially designed to compensate the reactive power and decrease the harmonic currents occurred on the side of load from the grid, by injecting the current having same magnitude but opposite in the phase direction of the harmonic current [7].

\*        Assistant Professor, Department of Electrical Engineering, Mehran University of Engineering & Technology, Jamshoro.
\*\*       Professor, Department of Electrical Engineering, Mehran University of Engineering & Technology, Jamshoro.
\*\*\*      Professor, Department of Computer Systems Engineering, Mehran University of Engineering & Technology, Jamshoro.

In the literature a number of methods has been emphasized for identification of reference current [8-13]. For identification of disturbing current the instantaneous reactive power method has been used in this paper. For controlling the output currents of the converter, to follow the reference currents, hysteresis current controller has been used in [14-19]. The implementation of hysteresis current controller is simple and its performance is good. Therefore hysteresis control technique has been used for generating the switching pulses for voltage source converter.

## 2. STRUCTURE OF HAPF

The simulink model of HAPF control is shown in Fig. 1. It is based on:

- A three-phase three-wire power circuit.
- Non-linear load based on uncontrolled rectifier.
- Reference current estimator block based on pq theory.
- Pulse generation block, based on hysteresis current control technique.
- The HAPF consists of a VSC (Voltage Source Converter) and capacitor.
- dc bus voltage controller block

## 3. ESTIMATION OF COMPENSATION CURRENT REFERENCE

For the control of HAPF the reference current can be determined by calculation of instantaneous power components of the pq theory in a stationary $\alpha\beta$ frame [20-21].

In the sequence of r,y,b, the axis planes for r,y,b are fastened but apart each other by $2\pi/3$. The direction and amplitude of instantaneous space vectors $e_\alpha$ and $i_\alpha$ varies with time when they are set on the $\alpha$-axis. The space vector transformation equations into $\alpha,\beta$ coordinates are stated as follows:

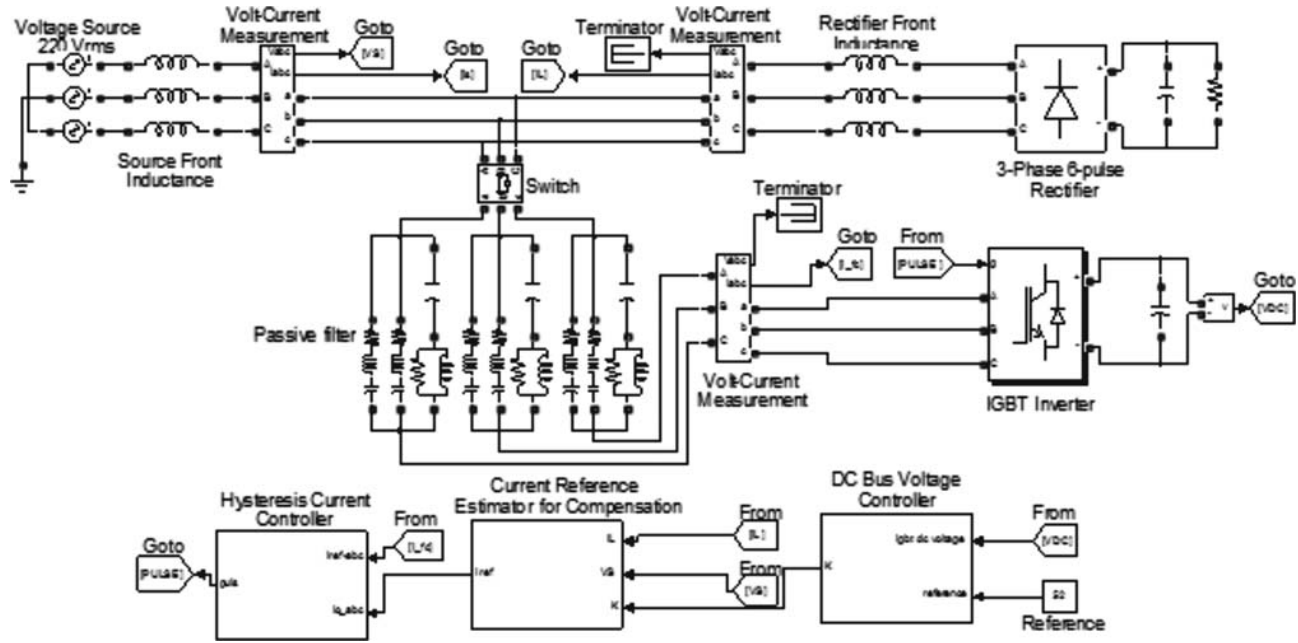

*FIG. 1. GENERAL STRUCTURE OF HYBRID ACTIVE POWER FILTER AND ITS CONTROL*

Design of Three-Phase Hybrid Active Power Filter for Compensating the Harmonic Currents of Three-Phase System

$$\begin{bmatrix} e_\alpha \\ e_\beta \end{bmatrix} = H \begin{bmatrix} e_r \\ e_y \\ e_b \end{bmatrix} \quad (1)$$

$$\begin{bmatrix} i_\alpha \\ i_\beta \end{bmatrix} = H \begin{bmatrix} i_r \\ i_y \\ i_b \end{bmatrix} \quad (2)$$

Where

$$H = \sqrt{\frac{2}{3}} \begin{pmatrix} 1 & -\frac{1}{2} & -\frac{1}{2} \\ 0 & \frac{\sqrt{3}}{2} & -\frac{\sqrt{3}}{2} \end{pmatrix}$$

The Equations (1-2) show that $\alpha$ and $\beta$ coordinates are orthogonal to each other. They also indicate that $e_\alpha$ and $i_\alpha$ are on the $\alpha$-axis while $e_\beta$ and $i_\beta$ are on the $\beta$-axis. In 3-$\Phi$ circuit the equation of instantaneous real power can be computed as:

$$p = e_r i_r + e_y i_y + e_b i_b = e_\alpha i_\beta + e_\beta i_\beta + e_\beta i_\beta \quad (3)$$

The instantaneous reactive power can be computed as:

$$q = e_\alpha i_\beta + e_\beta i_\alpha \quad (4)$$

From Equations (3-4), the instantaneous powers are recomputed as in Equation (5)

$$\begin{bmatrix} p \\ q \end{bmatrix} = K \begin{bmatrix} i_\alpha \\ i_\beta \end{bmatrix} \quad (5)$$

Where

$$K = \begin{pmatrix} e_\alpha & e_\beta \\ -e_\beta & e_\alpha \end{pmatrix}$$

Based on theoritical approach, the thee-phase voltage $v_r$, $v_y$ and $v_b$ and currents of the nonlinear load can be transformed in to $\alpha\beta$ frame as:

$$\begin{pmatrix} e_\alpha \\ e_\beta \end{pmatrix} = H \begin{pmatrix} v_r \\ v_y \\ v_b \end{pmatrix} \quad (6)$$

and

$$\begin{pmatrix} i_{L\alpha} \\ i_{L\beta} \end{pmatrix} = H \begin{pmatrix} i_{Lr} \\ i_{Ly} \\ i_{Lb} \end{pmatrix} \quad (7)$$

The instantaneous powers ($p_L$ and $q_L$) of the load can be computed as:

$$\begin{pmatrix} p_L \\ q_L \end{pmatrix} = K \begin{pmatrix} i_{L\alpha} \\ i_{L\beta} \end{pmatrix} \quad (8)$$

Equation (8) is rearranged as:

$$\begin{pmatrix} i_{L\alpha} \\ i_{L\beta} \end{pmatrix} = K^{-1} \begin{pmatrix} p_L \\ q_L \end{pmatrix} \quad (9)$$

The $p_L$ is divided in to average power ($\bar{p}_L$) and oscillating power ($\tilde{p}_L$). Similarly $q_L$ is divided in to average power ($\bar{q}_L$) and oscillating power ($\tilde{q}_L$) as in equation (10).

$$p_L = \bar{p}_L + \tilde{p}_L \text{ and } q_L = \bar{q}_L + \tilde{q}_L \quad (10)$$

From Equation (9) the $\alpha$-phase load current $i_{L\alpha}$ can be computed as follows:

$$i_{L\alpha} = \frac{e_\alpha}{e_\alpha^2 + e_\beta^2} \bar{p}_L + \frac{-e_\beta}{e_\alpha^2 + e_\beta^2} \bar{q}_L + \frac{e_\alpha}{e_\alpha^2 + e_\beta^2} \tilde{p}_L + \frac{-e_\beta}{e_\alpha^2 + e_\beta^2} \tilde{q}_L \quad (11)$$

From Equation(11), it can be observed that in order to cancel out harmonic distortion in power system line, the second, third and fourth term of Equation (11) must be compensated by the HAPF.

The reference compensating currents can be expressed as follows:

$$\begin{pmatrix} i_r^* \\ i_y^* \\ i_b^* \end{pmatrix} = \sqrt{\frac{2}{3}} \begin{pmatrix} 1 & 0 \\ -\frac{1}{2} & \sqrt{3}/2 \\ -\frac{1}{2} & -\sqrt{3}/2 \end{pmatrix} K^{-1} \begin{pmatrix} p^* + p_{ave} \\ q^* \end{pmatrix} \quad (12)$$

where $p_{ave}$ is the instantaneous value of the true power related to the HAPF loss and $p^*$ and $q^*$ are given by:

$$p^* = -\tilde{p}_L$$
$$q^* = -q_L \quad (13)$$

## 4. VOLTAGE CONTROLLER OF DC-BUS

The main function of DBVC (Voltage Controller of DC Bus) is to control power supply to the insulated gate bipolar transistor converter. To avoid using external power such as DC power supply, Batteries, and many more, the charging and discharging phenomenon of capacitor element is manipulated. The charging and discharging phenomenon of capacitor is controlled by the DBVC for maintaining the DC voltage level at constant and the transient response reached the stability. There are several approaches that can be implemented for the DBVC such as using the PI controller but in this paper, the approach used is based on [22]. Fig. 2 shows the voltage controller of DC bus. Voltage controller of DC-bus regulates the voltage of the capacitor on DC side of the converter and compensates losses of the converter. The voltage of the capacitor is measured and then it is compared to constant value. The resultant difference is then fed into gain. In this way, HAPF regulates and builds up the DC voltage of capacitor without using any external source.

## 5. HYSTERESIS CONTROL TECHNIQUE

With help of hysteresis control technique gating signals for voltage source converter are generated. For controlling the output currents of the HAPF various control techniques have been used. Among these techniques hysteresis current control techniques is the most popular technique for the application of the HAPF. In this technique, as shown in Fig. 3, the actual compensation current reference is compared with pre-described band of hysteresis around reference current. The resultant current will feed into the hysteresis band comparator which generates the switching signals for VSC. The switching pulses are generated when actual compensation current reference crosses the upper and lower tolerance bands. If actual compensation current reference is within a tolerance band then no switching pulses are generated.

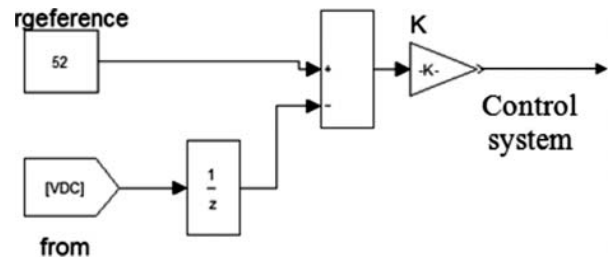

*FIG. 2. VOLTAGE CONTROLLER OF DC BUS*

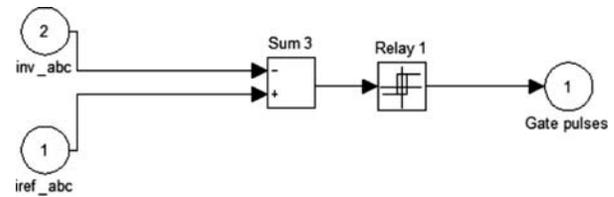

*FIG. 3. HYSTERESIS BAND CURRENT CONTROLLER*

Design of Three-Phase Hybrid Active Power Filter for Compensating the Harmonic Currents of Three-Phase System

## 6. SIMULATION RESULTS

The circuit parameters used in simulation are presented in Table 1.

Simulations were performed to show the usefulness of the HAPF which actually compensates the harmonic currents of the source current effectively.

A nonlinear load, consisting of a 3-Φ uncontrolled rectifier and adjustable speed drives, is fed by sinusoidal and symmetrical mains phase voltages (220 Vrms, 50 Hz).

Figs. 4-5 show the source currents before compensation and its harmonics spectrum, respectively.

From Fig. 5 it is clear that the THD of the source current is 20.77%. Therefore without using HAPF the THD is out of range set by the IEEE 519 standards.

**TABLE 1. SIMULATION PARAMETERS [23]**

| Supply Voltage | Vs=220Vrms |
|---|---|
| Supply/Line Inductance | $L_s$ = 0.0016 Henery |
| Rectifier Front-End Inductance | $L_L$ = 0.023 Henery |
| Capacitance of the Load | $C_L$ = 50 Micro Farad |
| Resistance of the Load | $R_L$ = 78 Ohm |
| Capacitance of Inverter dc-Bus | $C_{dc}$ = 4500 Micro Farad |

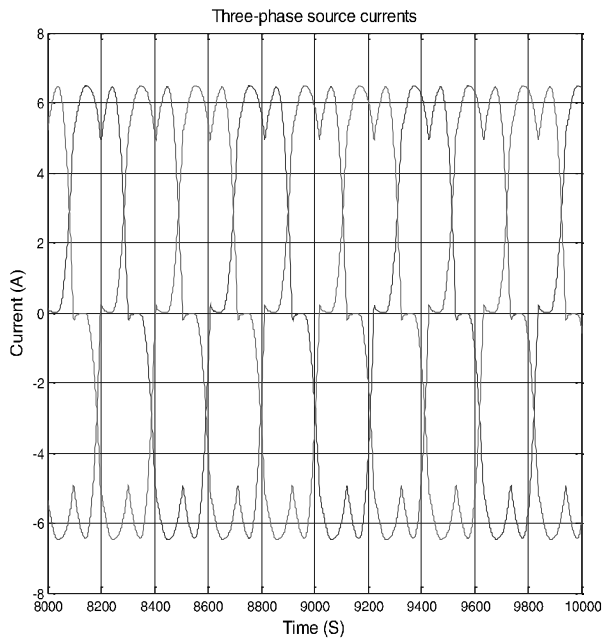

FIG. 4. SUPPLY CURRENTS WAVEFORMS BEFORE FILTERING

In order to follow the power quality standards the total harmonic distortion should be less than 5%. This can be realized by designing the HAPF. The parameters of the proposed PFs for HAPF have been shown in Table 2.

The simulation results of the proposed HAPF have been presented in Figs. 6-9. After compensation the source currents and source voltages are depicted in Figs. 6-7 respectively. The source currents are near to sinusoidal and its THD is reduced from 20.77-1.97% as depicted in Fig. 8. Thus power factor is improved. In Fig. 9, the output current of HAPF tracks the reference current closely.

## 7. CONCLUSION

The quality of the waveform is based on the value of total harmonic distortion. Therefore, THD factor of the source current is analyzed in time and frequency domains. From the results it is evident that the HAPF effectively reduces the THD produced by nonlinear load. The main objective of this research work has been accomplished. The total distortion of the supply current has been decreased at a high level of expectation from 20.77 to 1.97% in the simulation, which is an achievement to meet the IEEE 519 recommended harmonic standard.

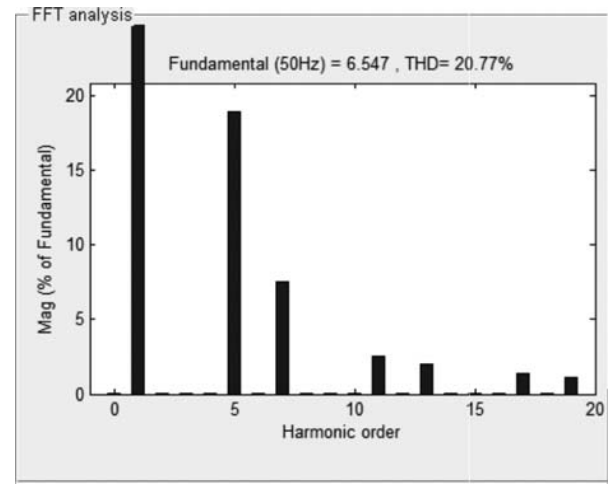

FIG. 5 SUPPLY CURRENT SPECTRUM BEFORE FILTERING

**TABLE.2. VALUES OF DESIGNED FILTERS**

| C (F) | L (H) | R (Ω) |
|---|---|---|
| $C_{5th}$ = 20e-6 | $L_{5th}$ = 0.0199 | $R_{5th}$ = 0.629 |
| $C_{7th}$ = 10e-6 | $L_{7th}$ = 0.0204 | $R_{7th}$ = 0.902 |
| $C_{HP}$ = 3.25e-6 | $L_{HP}$ = 0.025 | $R_{HP}$ = 260 |

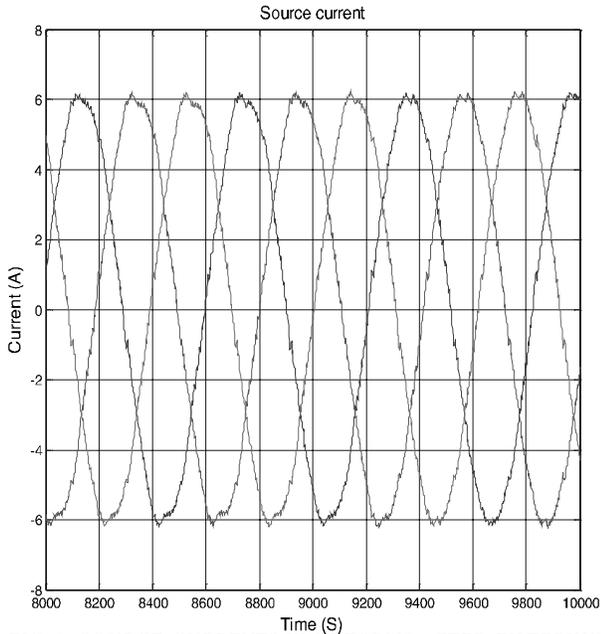

FIG. 6. SOURCE CURRENTS WAVEFORMS AFTER FILTERING

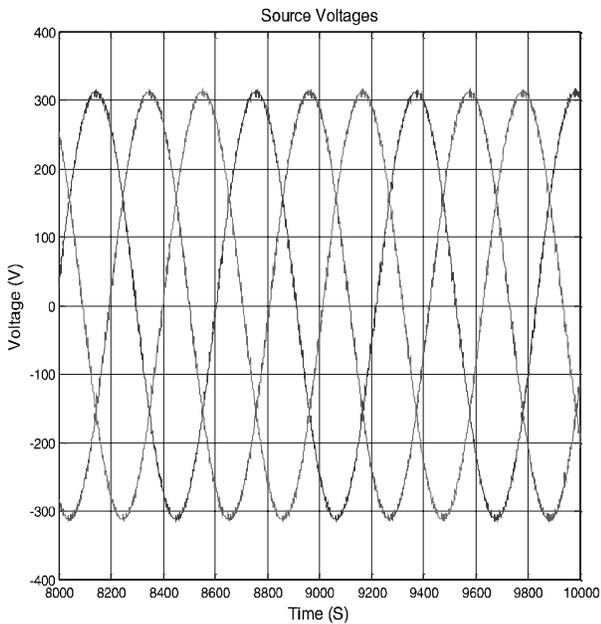

FIG. 7. SOURCE VOLTAGES WAVEFORMS AFTER FILTERING

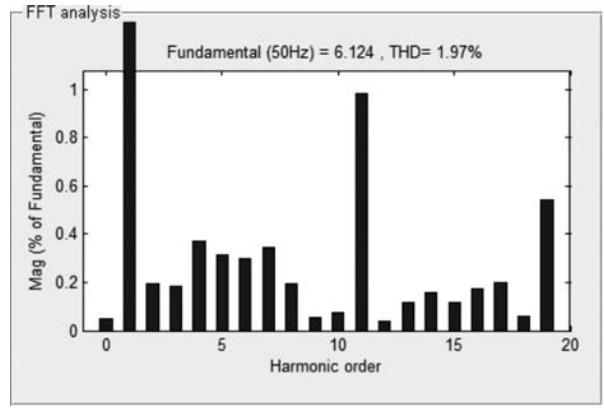

FIG. 8. SPECTRUM OF SOURCE CURRENT AFTER FALTERING

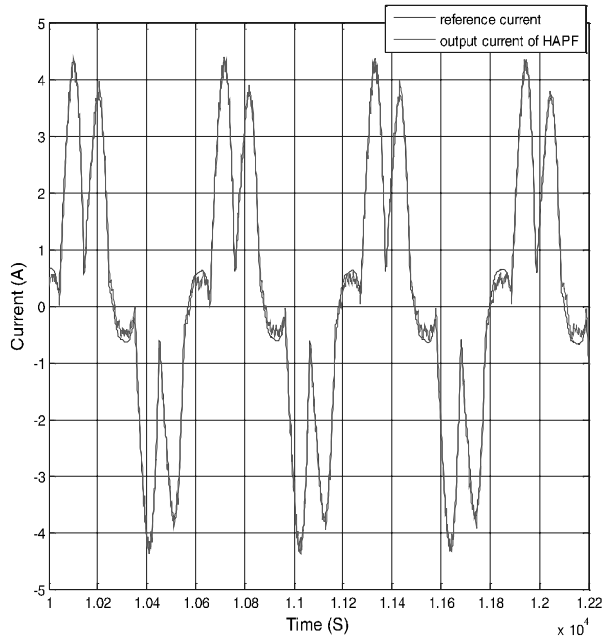

FIG. 9. HYBRID ACTIVE FILTER OUTPUT CURRENT AND REFERENCE CURRENT

## ACKNOWLEDGMENT

Authors acknowledges with thanks the higher authotiries and Department of Electrical Engineering, Mehran University of Engineering & Technology, Jamshoro, Pakistan, for providing moral support and necessary facilities to complete this research work.